\documentclass[]{interact}

\usepackage[utf8]{inputenc}
\usepackage{hyperref}

\usepackage{siunitx}
\DeclareSIUnit\bar{bar}
\usepackage[dvipsnames]{xcolor}
\usepackage{soul}

\newcommand{\revadd}[1]{#1}
\newcommand{\revrem}[1]{}

\usepackage[english]{babel}

\usepackage[caption=false]{subfig}

\usepackage[numbers,sort&compress]{natbib}
\bibpunct[, ]{[}{]}{,}{n}{,}{,}
\renewcommand\bibfont{\fontsize{10}{12}\selectfont}
\makeatletter
\def\NAT@def@citea{\def\@citea{\NAT@separator}}
\makeatother

\begin{document}

\articletype{REVIEW ARTICLE}

\title{High-throughput compfutational screening of nanoporous materials in targeted applications}

\author{
\name{Emmanuel Ren\textsuperscript{a,b}, Philippe Guilbaud\textsuperscript{b} and Fran\c{c}ois-Xavier Coudert\textsuperscript{a}\thanks{CONTACT Fran\c{c}ois-Xavier Coudert. Email: fx.coudert@chimieparistech.psl.eu}}
\affil{\textsuperscript{a} Chimie ParisTech, PSL University, CNRS, Institut de Recherche de Chimie Paris, 75005 Paris, France; \textsuperscript{b} CEA, DES, ISEC, DMRC, Univ. Montpellier, Marcoule, France}
}

\maketitle

\begin{abstract}
Due to their chemical and structural diversity, nanoporous materials can be used in a wide variety of applications, including fluid separation, gas storage, heterogeneous catalysis, drug delivery, etc. Given the large and rapidly increasing number of known nanoporous materials, and the even bigger number of hypothetical structures, computational screening is an efficient method to find the current best-performing materials and to guide the design of future materials. This review highlights the potential of high-throughput computational screenings in various applications. The achievements and the challenges associated to the screening of several material properties are discussed to give a broader perspective on the future of the field.
\end{abstract}

\clearpage


\section{Introduction}

Nanoporous materials are characterised by a high internal surface area on which a large number of molecules can physically or chemically adsorb on. They can thus be used in various key sectors of the industry, such as gas separation and capture,\cite{Li_2009} storage,\cite{Morris_2008} heterogeneous catalysis\cite{Bell_2003,Singh_2019} or drug delivery.\cite{Della_Rocca_2011,Bernini_2014} Among notable examples we can cite H$_2$ and CH$_4$ purification and storage, CO$_2$ capture, CO removal for fuel cell technology, desulfurisation of transportation fuels, and other technologies for meeting increasingly higher environmental standards. Moreover, nanoporous materials can have different chemical natures (inorganic, organic, or hybrid) and porosity (macroporous, mesoporous, or microporous). This opens up a large space of possible properties to explore and to find the most suitable structure for each specific application.

\begin{figure}[ht]
\centering
  \includegraphics[width=0.6\linewidth]{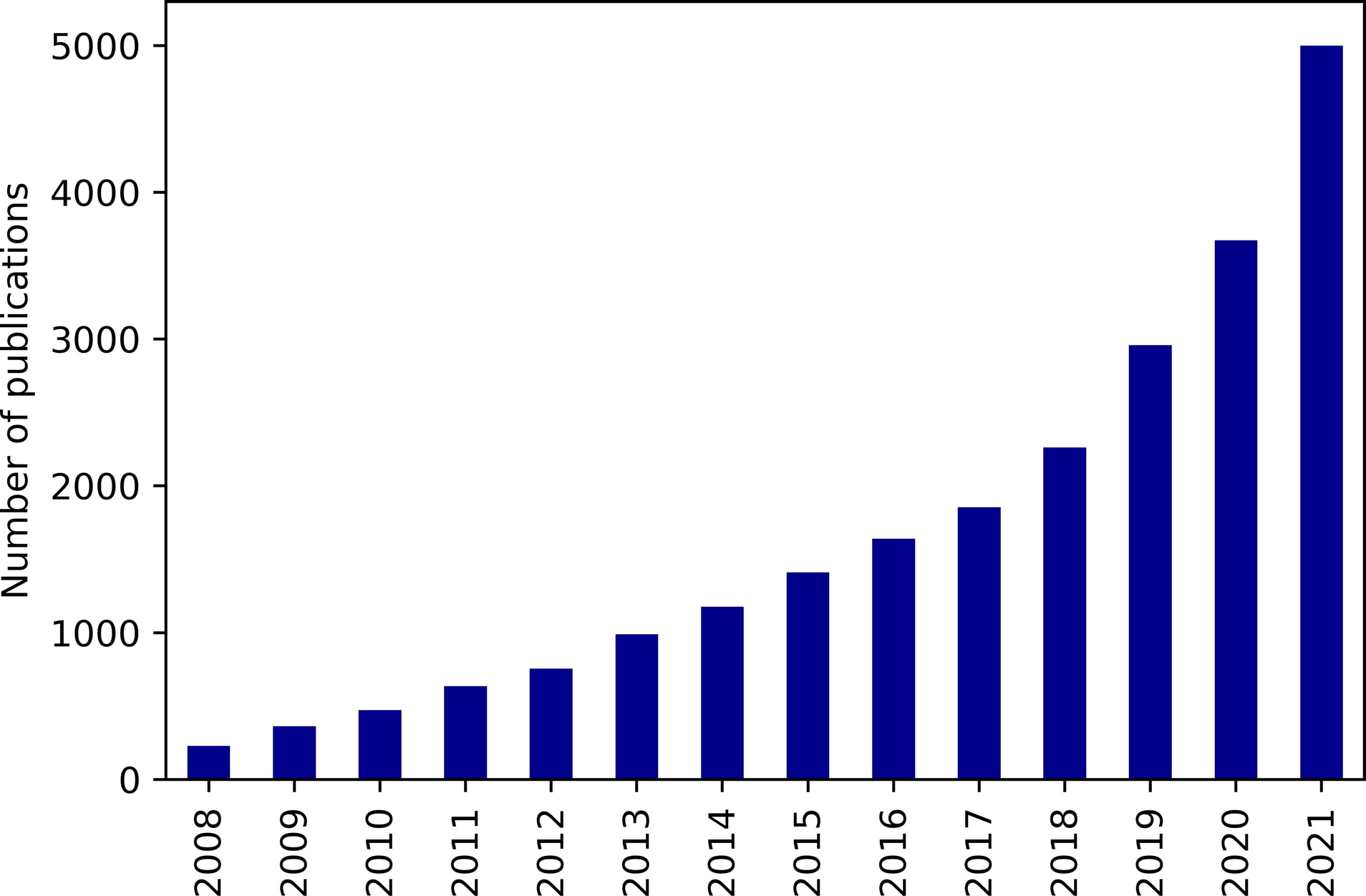}
  \caption{Number of scientific publications per year for computational screening of nanoporous materials, from 2008 to 2021, highlighting the acceleration of research in this area in the past decade (data from Scopus). \revadd{Over the same period of time, the total number of articles published in chemistry and materials sciences has grown from 356,000 to 602,000.}}
  \label{fgr:research_growth}
\end{figure}

Nanoporous materials can be used in a very wide range of applications, but systematically identifying the best material may seem like searching for a needle in a haystack. In fact, hundreds of thousands structures have been synthesised and possibly millions of materials are yet to be studied. A purely experimental approach, in addition to be expensive and time-consuming, would never be exhaustive to screen all these structurally and chemically diverse materials. Beyond this experimental limitation, large-scale computational screening studies can enable a more in-depth exploration of the existing materials, as well as generate novel hypothetical structures with potentially better performance. Even if the idea of this thorough exploration and the required databases of computationally-generated or experimentally-sourced structures were known for a very long time,\cite{PDB1971, Grazulis2009, Groom2016} research interest on computational screening applied to nanoporous materials has just experienced a rapid growth in the last decade (see Figure \ref{fgr:research_growth}). Several factors can explain this recent expansion: 1) the emergence of open databases of material structures and properties has opened the access for a growing number of scientists;\cite{Coudert2019, dePablo2014, dePablo2019, Jain2013, Jain2016} 2) the advances in the \emph{in silico} construction of hypothetical nanoporous materials have created new datasets to explore;\cite{Foster2004, Wilmer2013, Boyd2016} \revadd{3) efficiently implemented open-source software have granted access to simulation tools for a much larger research community;\cite{Willems2012,dubbeldam2016} 4) increasingly efficient supercomputers are now more and more available;\cite{Lim2015}} \revadd{5}) text and data mining have generated new databases of unreported properties from existing literature;\cite{Tshitoyan2019, Court2020} \revadd{6}) and the size of screenable databases have been increased by several orders of magnitude thanks to artificial intelligence techniques.\cite{Butler2018, Kim2017, Borboudakis2017, Chibani2020}

Given the aforementioned scientific advances, computational screening, that was commonly used on small series of materials, began to be used on larger databases to identify top performing candidates, to better understand the main explanatory factors at the origin of the performance and to objectively set theoretical performance limits for a given application. Borrowing some techniques from the new field of data science, screening techniques are now applied to predict key performance indicators.
These figures of merit are related to a variety of material properties such as electronic structure,\cite{Hachmann2011, Davies2016, Zhang2019} chemical and catalytic activity,\cite{Singh2015, Greeley2006, Back2020} thermal properties,\cite{Toher2014, Sarikurt2020, Ducamp_2021} mechanical properties,\cite{Chibani2019, Gaillac2020} transport and thermodynamic properties for adsorption.\cite{Watanabe2012, Kim2012, Han2012,Wilmer_2011}

The present work is by no mean an exhaustive review of all the works on the subject, but it aims at giving nonspecialist readers a high-level overview of the potential of computational screening in a large variety of applications, and of the diversity of the different approaches used in this field of research.
First, a brief survey of the development of materials databases and screening methodologies is given along with some examples illustrating the major milestones. Then, the thermodynamic properties linked to the adsorption processes are thoroughly reviewed; before moving to kinetic effects, looking at the prediction of transport properties. Finally, other aspects that differ from the adsorption process such as the computational screening of mechanical, thermal and catalytic properties are described at the end. We conclude by outlining some of the perspectives of the field.

\section{Screening methodologies}

\begin{figure*}[t]
\centering
  \includegraphics[width=\textwidth]{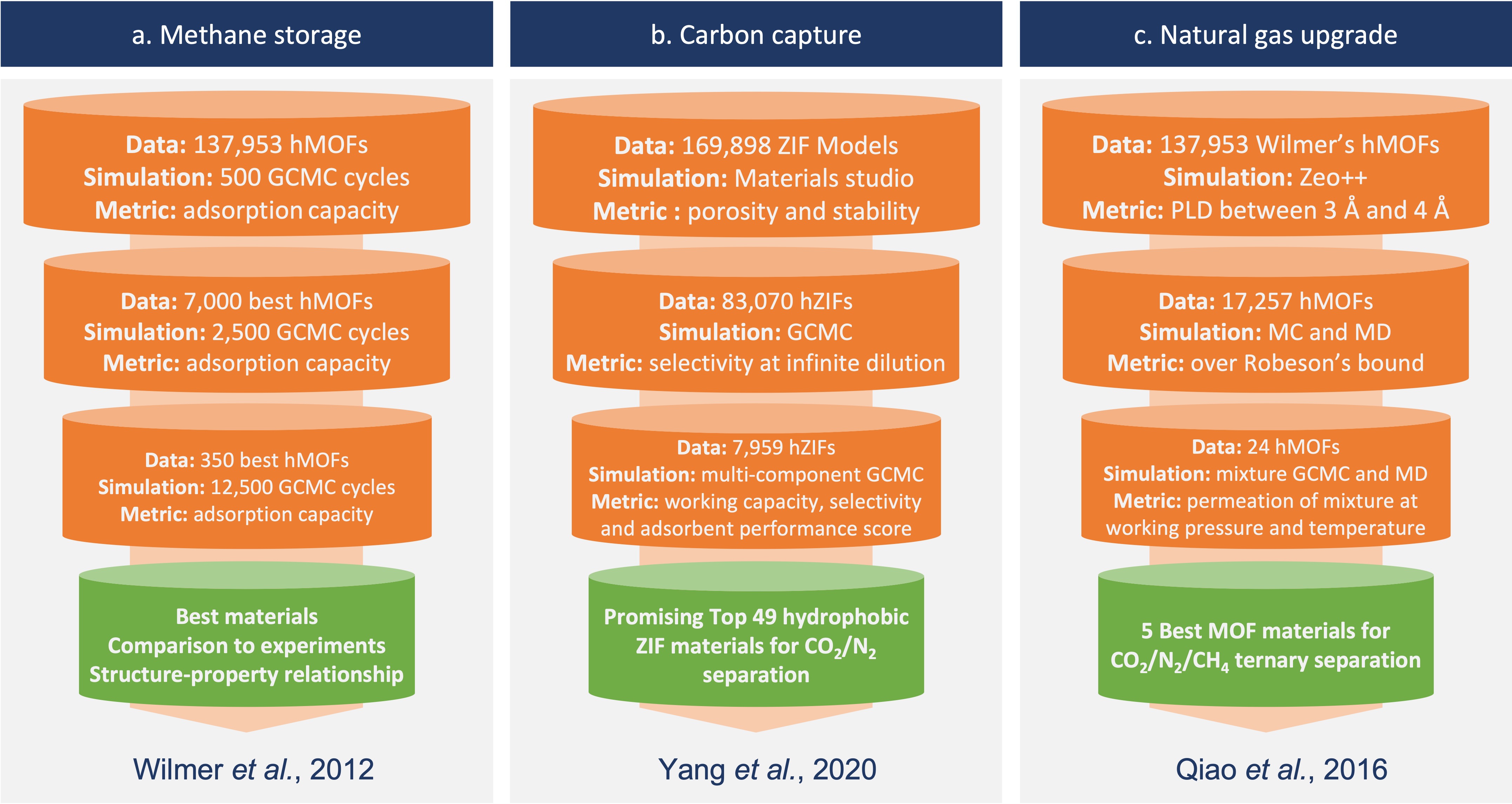}
  \caption{Simplified representation of typical funnel-type screening procedures, exemplified on three different applications from the published literature. (a) Wilmer \emph{et al.}\cite{Wilmer_2011} used a series of bi-component Grand Canonical Monte Carlo (GCMC) calculations at different levels of complexity to screen a large dataset of hypothetical MOFs for methane storage application. (b) Yang \emph{et al.}\cite{Yang_2020} used simulations at infinite dilution to pre-screen the dataset before using computationally demanding simulations and multiple metrics to find the most promising ZIFs for carbon capture. (c) In Qiao \emph{et al.}\cite{Qiao_2016}, transport properties were screened along standard adsorption properties to find the best materials for the targeted CO$_2$/N$_2$/CH$_4$ ternary separation; similarly, cheaper calculations at infinite dilution were carried out in a first step, before using more expensive calculations at working pressure and temperature.}
  \label{fgr:screening}
\end{figure*}

\subsection{Nanoporous material databases}

Before building any screening strategy or performing any computational screening, one needs to generate a set of files describing the atomic structure of the materials. Nanoporous materials can \revadd{have different degrees of crystallinity from perfectly crystalline to completely amorphous}. Most of the computational work is focused on crystalline structures, since the atoms are well-described within a periodic framework\revadd{, which enables faster simulations}. \revadd{The presence of defects are also usually neglected, which could explain some of the discrepancies between simulations and experiments. And amorphous materials are described by thousands of atomic positions in order to grasp their intrinsic non-periodicity.}\cite{Thyagarajan_2020} One can distinguish roughly four main classes of crystalline nanoporous materials: the inorganic zeolites (\emph{e.g.} aluminosilicates, aluminophosphates), the porous polymer networks, the covalent organic frameworks (COFs) and the metal--organic frameworks (containing the zeolitic imidazolate frameworks \emph{ie.} ZIFs and others). This diversity of nanoporous materials offer a wide range of potential candidates for any targeted applications.

The International Zeolite Association \revadd{(IZA) }gave a standardised set of 244 zeolites \revadd{(in their idealized all-silica form)} that can be used for screening purposes. To generate a dataset of structures, existing experimental database like the Cambridge Structural Database can be exploited. However, the raw structures determined experimentally by X-ray cannot be used directly as is. To obtain a computation-ready dataset, Chung \emph{et al.} used algorithmic cleaning procedures to build the publicly available Computation-Ready Experimental MOF (CoRE MOF) database.\cite{Chung_2014, Chung2019} CoRE MOF 2019 contains about 14,000 MOF structures, which is the biggest experimental database. Similar approach applied to organic frameworks led to the construction of a set of 187 COFs with disorder-free and solvent-free structures.\revadd{\cite{Tong_2017,Ongari_2019}}

These experiment-based databases can already be used in computational screenings to retrieve valuable information, but unknown structures that are yet to be discovered are not represented. To overcome the limits and biases of experimental synthesis, artificial ways of generating nanoporous material datasets can be used, which proved to be extremely efficient. The first \emph{in silico} generated database of about 130,000 MOFs used a recursion-based assembly (or tinkertoy-like) algorithm to combine 102 building blocks.\cite{Wilmer_2011} Martin and Haranczyk then proposed a topology-specific structure assembly algorithm that leverage the topological information of the structures.\cite{martin2014construction} Inspired by this algorithm, topology-based databases emerged a few years later with the set of 13,000 MOF structures generated using the Topologically Based Crystal Constructor (ToBaCCo) algorithm \revadd{developed by Colon, G{\'{o}}mez-Gualdr{\'{o}}n and Snurr}.\cite{Colon2017}
Later, Boyd and Woo proposed another topology-based algorithm using a graph theoretical approach and generated a 300,000 structures database (BW-DB) based on 46 different network topologies.\cite{Boyd_2016}
Similar approaches are used for other classes of materials, Deem and coworkers proposed a dataset of nearly 2.6 million hypothetical zeolite structures.\cite{Earl_2006,Deem_2009,Pophale_2011}
However, one could wonder if these hypothetical structures are synthesisable and can remain stable under operational conditions (\emph{e.g.} thermal, mechanical, radioactive constrain\revadd{t}s). \revadd{To discuss their synthetic likelihood, }Anderson and G{\'{o}}mez-Gualdr{\'{o}}n computed the free energies of 8,500 hypothetical structures \revadd{and compared} them to experimentally observed MOF structures.\cite{Anderson_2020} \revadd{This type of prediction can be very useful as it enables to gauge the relative stability of each materials and to only consider the stable structures. Other types of materials have been explored, Turcani \emph{et al.} published 60,000 organic cage structures and used machine learning to predict their stability based on the shape persistence metric.\cite{Turcani2018}}

The Materials Genome Initiative, a 100 million dollar effort from the White House that aims to ``discover, develop, and deploy new materials twice as fast'', led to the creation of the ``Materials Project'', a centralised database containing all the above mentioned structures.\cite{kalil2011national,Mat_genome,Jain_2013}
The fast development of this nanoporous materials genome motivated Boyd \emph{et al.} to write a comprehensive review on all the initiatives on generating new data for computational analysis.\cite{Boyd_2017}

Yet, the sole increase in size of the databases is not enough. One needs to add diversity to have more general knowledge on the maximum performance and the explanatory features of such performance. Moreover, the diversity of structures ensure the quality of the predicted best materials for a given application.
To qualitatively or quantitatively assess the diversity of a database, inventive methodologies have been developed.
For instance, Martin, Smit and Haranczyk proposed a Voronoi hologram representation as a way of measuring similarities between structures to generate geometrically diverse subsets of a database.\cite{Martin_2011}
Moosavi \emph{et al.} made a comparative study of the diversity of three well-known databases CoRE MOF 2019,\cite{Chung2019} BW-DB\cite{Boyd_2016} and ToBaCCo\cite{G_mez_Gualdr_n_2016, Colon2017}  using geometrical and chemical descriptors to design a theoretical strategy for generating the most diverse set of materials.\cite{Moosavi_2020}
Another approach consists in searching for similarities instead of differences in the materials by studying topological patterns in the data.\cite{Lee_2017}
These investigations on the data structures give a solid ground to develop novel materials by objectively defining similarity, diversity and novelty. From the analysis gathered so far, one would need to radically change the approach by proposing materials with new chemistry, topology or mechanism (\emph{e.g.} flexibility) in order to significantly improve the diversity of the current databases.

\subsection{Evolution of screening methods}

In its early stage, computational screening has been used on small series of nanoporous materials to generate specific knowledge on some sub-classes of materials. These small-scale screenings combined with experiments helped faster identification of good performing candidates, but they failed to establish general rules of design or to explore the unknown. Larger-scale screenings overcame these limitations by trying to exhaustively cover the whole spectrum of nanoporous materials.

With the development of a nanoporous materials genome, several articles proposed methods to screen thousands of structures. Other challenges arose, such as the design of more efficient methods than the brute force screening or the analysis of big data. Two research groups \revrem{in Northwestern University} led by R. Snurr and J. Hupp began to address those questions, they used a ``funnel-like'' approach to efficiently screen about 130,000 hypothetical MOF structures.\cite{Wilmer_2011} To do so, they performed a first screening involving less steps of simulation on the whole dataset, then they extracted a subset of top performing structures to perform a second round with more simulation steps. This procedure is repeated until a few materials are selected by a final round of simulations with reasonable accuracy.
Similar ``funnel-like'' procedures have then been used in other field of applications as described in the Figure \ref{fgr:screening}. This type of screening saves precious computation time by balancing the complexity of the calculation with the amount of data to be screened. The most demanding simulations or experiments are only applied to the few most promising structures.
This method can rather efficiently identify top candidates, but it can't draw quantitative structure-property relationships (QSPR), beside facing scalability issues above a critical dataset size.

To overcome these new challenges, people are looking increasingly towards transferable models trained by a machine learning (ML) algorithm on a diverse and size-limited sub-sample. Ideally, such a model is transferable to potentially millions of structures and can provide valuable QSPR. For instance, Fernandez \emph{et al.}\cite{Fernandez_2013} used multiple linear regression analysis, decision tree regression, and nonlinear support-vector
machine models to extract QSPR and establish rules of designing well-performing MOFs for methane storage, while identifying promising structures. In this first work they only used geometrical descriptors to describe methane storage,\cite{Fernandez_2013} but realising the importance of chemical descriptors, they proposed the atomic property weighted radial distribution function as a powerful descriptor to predict CO$_2$ uptakes.\cite{Fernandez_2013_rdf}
More importantly, they proved that ML can be used as a pre-screening tool to avoid running time-costly simulations by correctly identifying around \SI{95}{\percent} of the top 1000 best performing materials. Recently, the same group used similar techniques to predict CO$_2$ working capacity as well as CO$_2$/H$_2$ selectivity in MOFs for precombustion carbon capture.\cite{Dureckova_2019}

\subsection{ML-assisted high-throughput screening}

\begin{figure*}[t]
\centering
  \includegraphics[width=\textwidth]{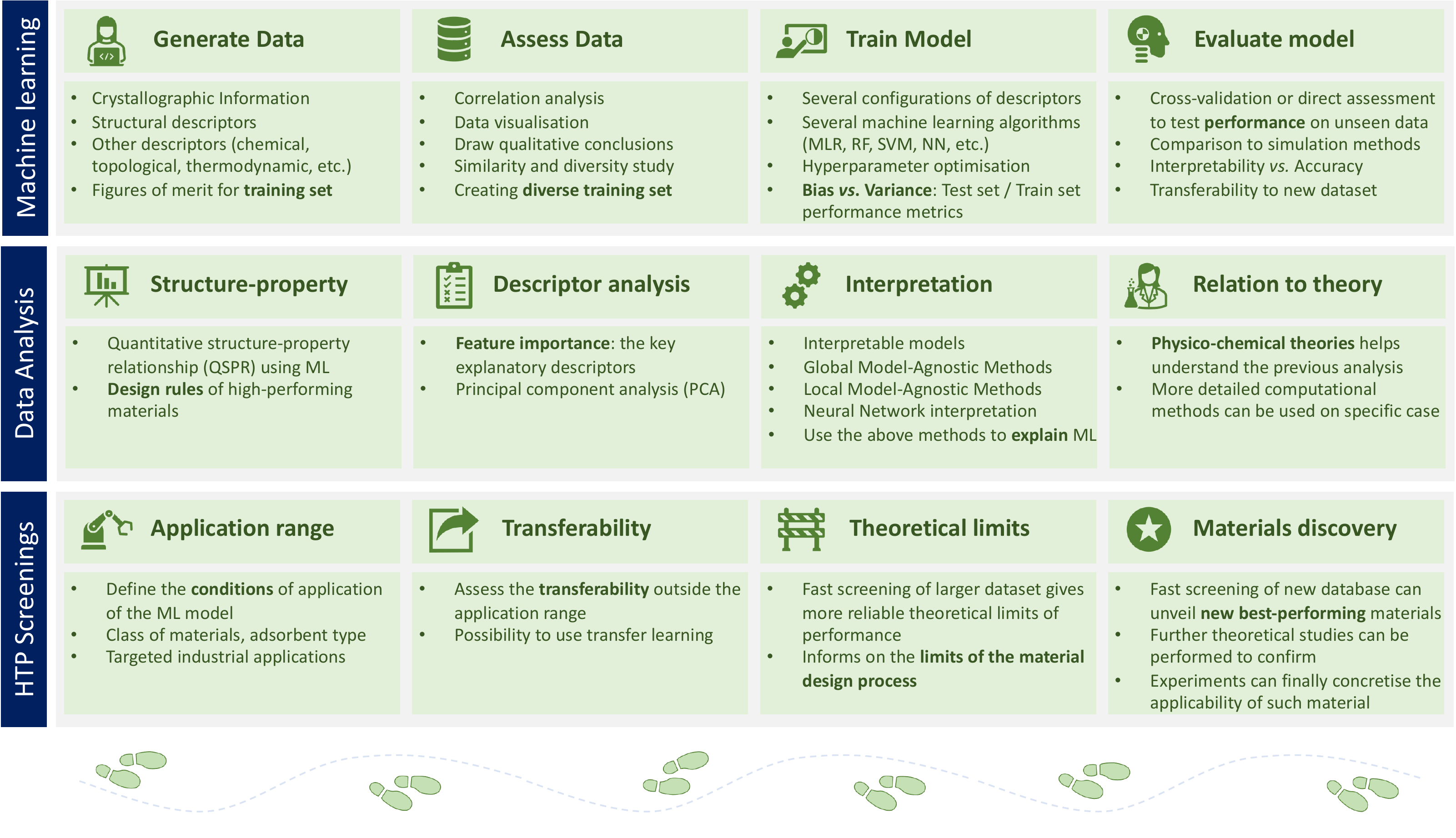}
  \caption{Schematic representation of the main subjects typically covered as part of an ML-assisted high-throughput screening procedures. First, one needs to train an machine learning (ML) model and analyse its performance on an independent subset of the data. Then, one can use the model to quantitatively extract structure--property relationships. Finally, once proven accurate, the model can be used on a larger scale to accelerate screening procedures.}
  \label{fgr:ML}
\end{figure*}

We saw the use of ML in the comprehension of the structure-property relationships, but it can also assist high-throughput screenings as illustrated in the Figure \ref{fgr:ML}. In an ML-assisted screening, one needs to consider first the type of algorithm and the features or descriptors. The descriptors exhaustively describe the physicochemical properties, while the ML algorithms set rules for learning patterns in the data. At the end, the ML model needs to be predictive while maintaining a high level of interpretability\revadd{\cite{Oviedo_2021}} and reproducibility\revadd{\cite{Coudert2017}}. To illustrate this approach, a few studies of such ML-assisted high-throughput screenings and their particular contributions to the field are presented below.

Regarding energy descriptors, different ones can be used alongside the most basic geometrical ones. For instance, Simon \emph{et al.} introduced the Voronoi energy, combined with structural descriptors they used them to predict Xe/Kr selectivity of over 600,000 structures using a random forest model.\cite{Simon2015}
Bucior \emph{et al.} also used an energy-based descriptor, the energy histogram, to predict the cryogenic storage capacity of hydrogen three times faster than traditional simulations.\cite{Bucior_2019}

Descriptors \revadd{based on the analysis of} data have also been studied and enable to find similarly performing materials. Based on advanced knowledge on mathematics and topology, Lee \emph{et al.} used a topological data analysis-based descriptor, called persistent homology and resembling barcodes, to screen a zeolite database for methane storage and carbon capture applications.\cite{Lee_2018} Later, Yongjin Lee led his group to propose an ML prediction method using the same pore geometry barcodes.\cite{Zhang2019} More recently, Moosavi \emph{et al.} built geometric landscapes, a representation for energy-structure-function maps based on geometric similarity, quantified by persistent homology.\cite{Moosavi_2020_esf}

To model the chemical behaviour of materials, one developed several chemical descriptors. In particular, Borboudakis \emph{et al.} introduced the chemical building block as a feature or descriptor of their ML models. In their study, they integrated all the models into a unified algorithm called ``Just Add Data'' and concluded that random forest and support vector machine were outperforming the other algorithms they tested.\cite{Borboudakis_2017} Recently, the same group continued on providing a universal (transferable on different materials) ML algorithm by using the type of atom instead of the previous building block description, which led to an increased performance on the prediction of methane and carbon dioxide adsorption capacities.\cite{Fanourgakis_2020} Anderson \emph{et al.} used the chemical building blocks of the MOF and the Lennard-Jones parameters of existing or ``alchemical'' adsorbates to train a neural network model for adsorption isotherms prediction. \cite{Anderson_2020_ML}

Through the scope of different types of descriptors, we introduced some ML-assisted approach to computational screenings. Figure \ref{fgr:ML} gives a higher-level view on how machine learning is practically applied. One can find a more comprehensive review on big-data science applied to porous materials written by Jablonka \emph{et al.}.\cite{jablonka2020big} The authors go through the selection of diverse data, the design of meaningful descriptors, ML algorithms, the best practices in the training process of an ML model, the measurement of its performance and the interpretation of the model to avoid the ``black box'' effect.

Beyond the reluctance to apply data science to fundamental sciences, one should not associate machine learning with the ``end of theory''; physicochemical theories can guide the development of the descriptors at the base of any ML models and the interpretation of these models is impossible without scientific insights. \revadd{The laws of physics are not explicitly included in an ML model, interpretability and explainability methods can help cover these flaws by identifying potential nonphysical behaviours, or confirming its consistency in describing known physical behaviours, or unveiling unexpected scientific insights.\cite{Oviedo_2021} If the model fails to meet some standards, further developments are needed for the descriptors to contain all relevant information, or to draw a more consistent relationship between the descriptors and the desired metric. Without a well-designed (containing all physical information) set of descriptors, an ML approach cannot make reliable predictions.} The recent developments presented here are confirming this close interplay between data science and theory.

\section{Thermodynamic properties of adsorption}

In its early development, computational screening was mainly used to predict thermodynamic properties in adsorption processes. Three main applications have been identified in the associated literature: gas storage (for energy or medical applications), gas separation (noble gas, hydrocarbons, carbon dioxide, etc.) and post-combustion CO$_2$ capture. These applications are closely linked to urgent environmental and energy issues that are yet to be solved. Screening can guide the development of better performing materials by shedding light upon unknown structure-property relationship, probes possible theoretical limitations (unreachable targets) and identifies potential candidates that need to be experimentally tested.

\subsection{Gas storage}

One can leverage the high surface density of the nanoporous materials, especially the MOFs, to stock in very low-density gas. In the field of energy storage or transportation, natural gas (mainly methane) or hydrogen are considered plausible alternative fuels to replace conventional ones for transport. The US Department of Energy (US DOE) recently financed research programs and set target for methane and hydrogen storage. Nanoporous materials could reduce energy, infrastructure and security cost due to the required compression and cooling. In this section, we are focusing on high-throughput screening for methane storage in nanoporous materials, before broadening the scope hydrogen and other perspectives.

\begin{figure}[ht]
\centering
  \includegraphics[width=0.5\linewidth]{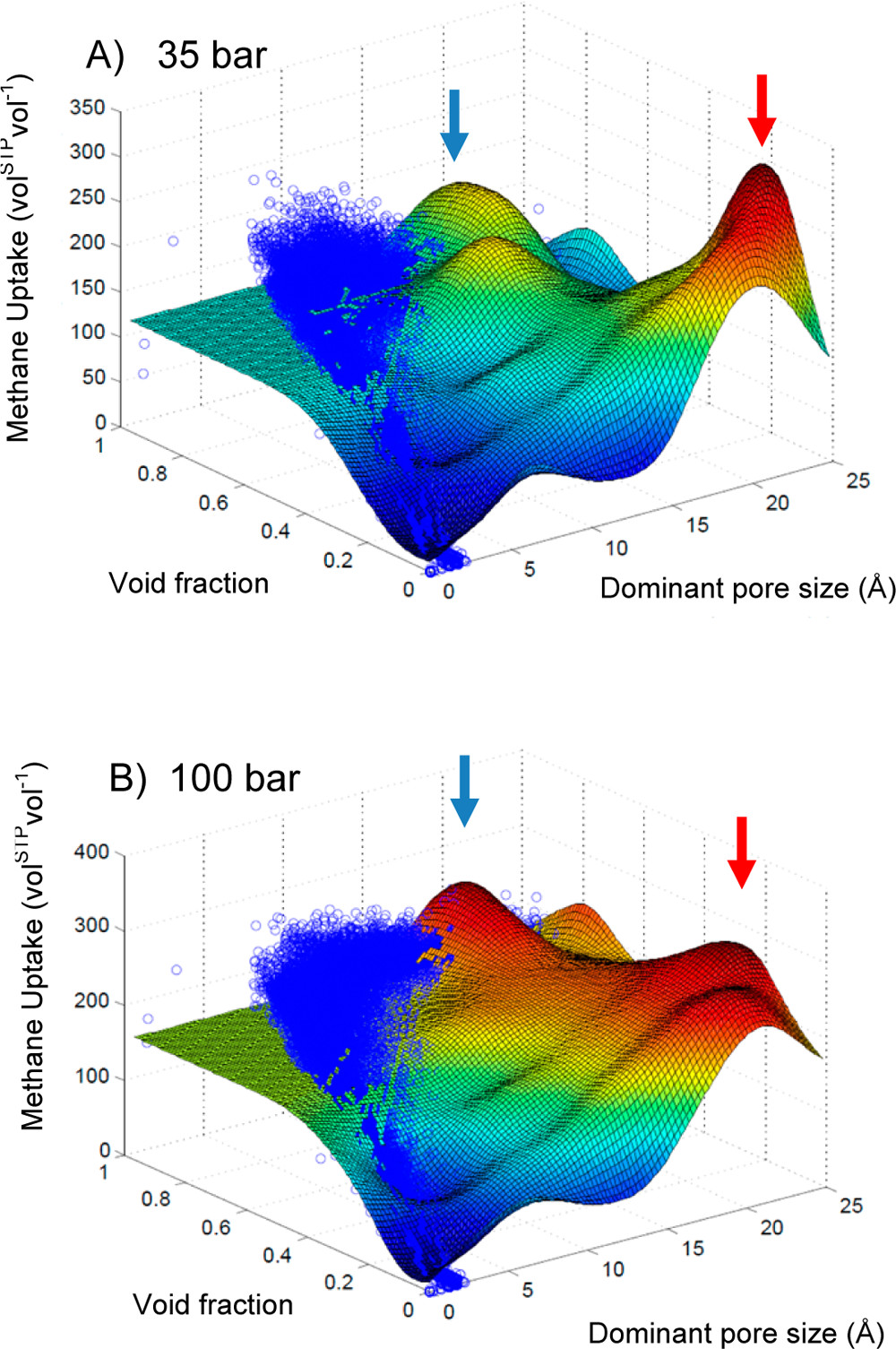}
  \caption{Two-dimensional response surfaces of the support vector machine (SVM) models trained by Fernandez \emph{et al.} for methane storage at (A) 35~bar and (B) 100~bar using void fraction and dominant pore size. The blue dots represent the GCMC simulated uptake values. The color of the surface represents the methane storage value, from blue (lowest values) to red (highest values). Blue and red arrows indicate maxima on the response surface. Reprinted with permission from Ref.~\citenum{Fernandez_2013}. Copyright 2013 American Chemical Society.}
  \label{fgr:Fernandez_2013}
\end{figure}

One of the pioneering works in computational screening was published in 2011 by Wilmer \emph{et al.}\cite{Wilmer_2011}. They performed a large-scale screening of 137,953 hypothetical MOF structures to estimate the methane storage capacity of each MOF at \SI{35}{\bar} and \SI{298}{\kelvin} based on the US DOE standards. Back then, the US DOE set a target methane capacity value of 180 vol$^\mathrm{\tiny STP}$ vol$^{-1}$ (which has since been achieved by several materials reported in the literature). In their large-scale analysis, Wilmer \emph{et al.} found over 300 hypothetical MOFs that meet the targeted requirements and the best one can store up to 267 vol$^\mathrm{\tiny STP}$ vol$^{-1}$, surpassing the state-of-the-art of the time. From their large dataset, a preliminary structure-property relationship analysis revealed that void fraction values of approximately 0.8 and gravimetric surface areas in a range 2500-3000 m$^2$g$^{-1}$ resulted in the highest methane capacities. Optimal pore size are also shown to be around the size of one or two methane molecule(s). Maximisation of gravimetric surface area was a common strategy in the MOF design for storage applications, but this study showed the existence of \revadd{an optimal range of surface area values}. Computational screenings can draw clear relationships between structural descriptors and performance. Later, a more quantitative relationship was drawn by Fernandez \emph{et al.} using ML models as illustrated on Figure \ref{fgr:Fernandez_2013}. \revadd{Beware not to over-interpret the relation given by the response surface, since the identified maxima do not always have a physical reality, especially where there is no training data in the area pointed by the red arrows. However, it highlights promising unexplored feature space and shows potential research directions.}

Since then new materials above the target have been found and the US DOE decided to set a higher target of 315 vol$^\mathrm{\tiny STP}$ vol$^{-1}$. Until now, this new target is not yet reached. This is why the recent developments have focused on assessing the feasibility of such a target by accelerating the screening methods so that more data can be screened, and by interpreting the QSPR models to extract important knowledge for the design of novel materials. For instance, G{\'{o}}mez-Gualdr{\'{o}}n \emph{et al.} showed that even by artificially quadrupling the Lennard-Jones interaction factor $\epsilon$ and by increasing the delivery temperature by \SI{100}{\kelvin}, the newly set target is only reached by a handful of MOFs.\cite{G_mez_Gualdr_n_2014} This study suggests the impossibility to reach the DOE target using a preconceived (experimentally or theoretically) material to store methane. However, this theoretical limitation can be overcome by increasing the surface density of sites with high affinity with methane and by increasing the delivery temperature.

Later, a larger-scale screening on methane storage was carried out by Simon \emph{et al.} on 650,000 experimental and hypothetical structures of zeolites, MOFs, and PPNs. This study confirmed that the classes of materials currently being investigated were unlikely to meet the new target. The authors suggested that it wasn't surprising since the target was based on economical arguments, while the screening is based on thermodynamic arguments.\cite{Simon2015_EES} This example illustrates the power of large scale screening to settle questions of physical feasibility \revadd{(if simulations are accurate)} and hence avoiding experimental efforts spent on impossible tasks.

More recently, a dataset containing trillions of hypothetical MOFs have been screened for methane storage.\cite{Lee_2021} Lee \emph{et al.} developed a methodology using machine learning combined with genetic algorithm to perform the largest screening until now. In addition to confirming most of the results (theoretical limits and QSPR) found by previous screenings, 96 MOFs were found to outperform the current world record. This study shows the scaling potential of ML-assisted screenings in handling ``Big data''.

Similarly computational high-throughput screenings have been applied to other storage applications such as hydrogen storage. Computational screenings showed that cryogenic storage of hydrogen can meet the DOE target of \SI{50}{\gram\per\liter}.\cite{G_mez_Gualdr_n_2016, Bobbitt_2016, Thornton_2017} Anderson \emph{et al.} performed a large scale screening based on neural networks to test out multiple pressure/temperature swing conditions to find that the maximal deliverable capacity cannot exceed \SI{62}{\gram\per\liter}.\cite{Anderson_2018} Compared to the density of liquid hydrogen (\SI{72}{\gram\per\liter}), this upper limit seems reasonable since the adsorbent material takes at least 10-20$\%$ of the tank. Here, we only showed some flagship results of the field. For a more detailed meta-analysis, Bobbitt and Snurr wrote a very complete review on computational high-throughput screening of MOFs for hydrogen storage.\cite{Bobbitt_2019}

\subsection{Gas separation}

As a representative example of what could be done in the field of gas separation, we are going to focus on Xe/Kr separation. These noble gases have multiple applications in the medical (\emph{e.g.} anaesthesia, painkiller),\cite{cullen1951anesthetic, holstrater2011intranasal} aeronautical\cite{Patterson_2002,Coxhill_2005} or lighting sectors,\cite{Jarman_1974,Tanaka_2019} just to cite a few. The industry more commonly uses cryogenic distillation to separate xenon and krypton from the ambient air, which requires a compression and cooling of the gas mixture. But this technology can laboriously be deployed in very security-sensitive cases such as the treatment of radioactive off-gases from nuclear plants. Nanoporous materials can be used as a safer, cheaper and less energy-intensive alternative. Computational screenings is an ideal tool to kick-start the development of this new technology by identifying rapidly the best candidates.

The first large-scale computational screening on Xe/Kr adsorption-based was performed by Sikora \emph{et al.} based on the same approach previously developed for methane storage by their group at the Northwestern University.\cite{Sikora2012} This study was based on the same {137,000} structures of hypothetical MOFs.\cite{Wilmer_2011} They calculated the Xe/Kr selectivity using Monte Carlo molecular simulations on the whole database by iteratively increasing the number of steps and selecting the best materials similar to the approach on Figure \ref{fgr:screening}. By analysing the relationships between pore sizes and selectivity, they confirmed a hypothesis from a smaller scale study that the pores should be between the size of 1 to 2 xenon molecules.\cite{Ryan_2010} Tube-like channel were also found to favour better selectivity. Moreover, they found that top performing materials could have selectivities around 500; but we can only conclude on the order of magnitude of the theoretical limitation of the Xe/Kr selectivity, considering the statistical uncertainty of the simulation.

\begin{figure}[ht]
\centering
  \includegraphics[width=0.7\linewidth]{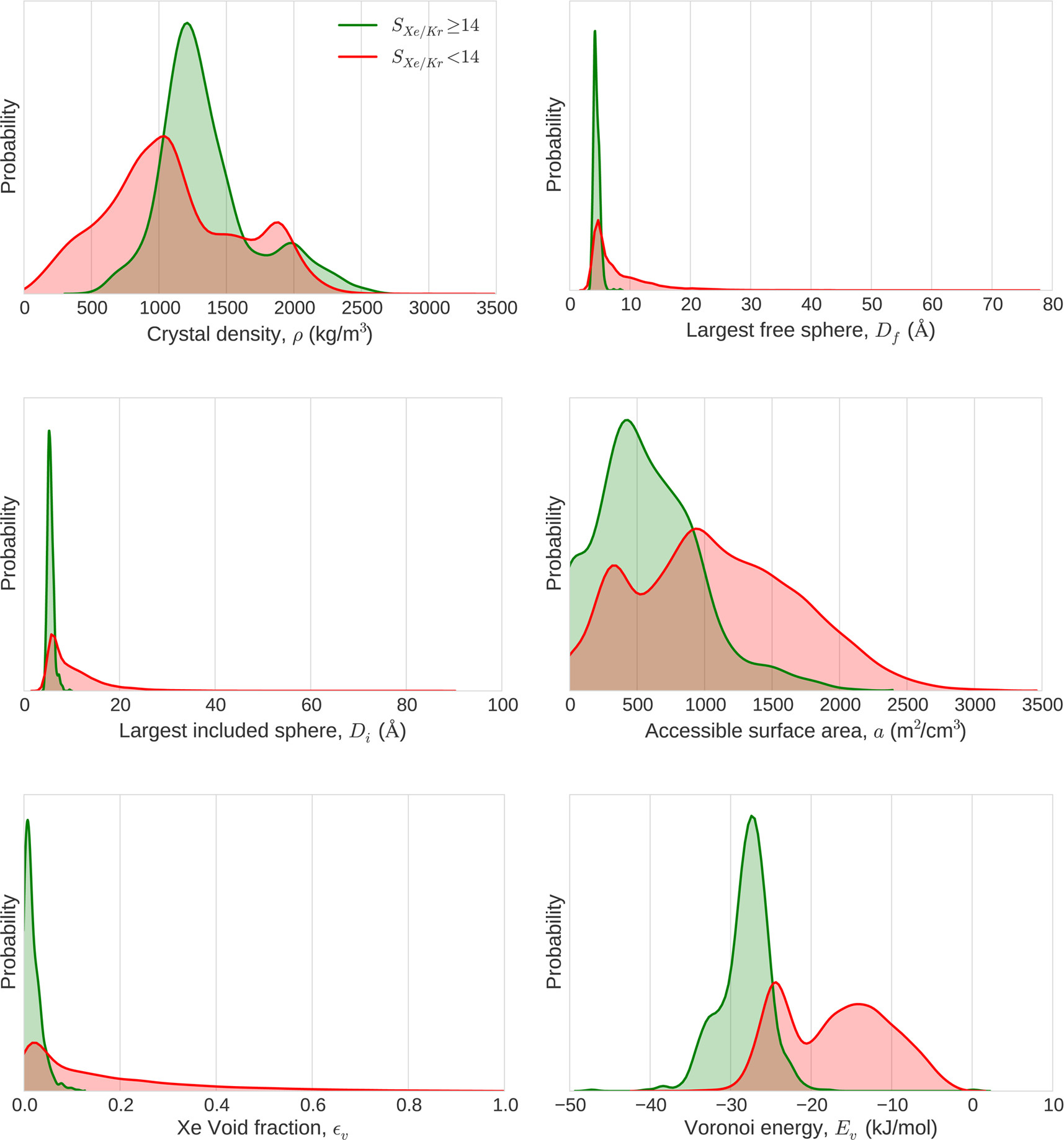}
  \caption{Statistical analysis of the adsorptive separation of xenon/krypton mixtures by nanoporous materials. The graphs represent the distributions of structural descriptors explored by highly selective (green) and poorly selective (red) materials separately. Reprinted with permission from Ref.~\citenum{Simon2015}. Copyright 2015 American Chemical Society.}
  \label{fgr:Simon2015}
\end{figure}

Seizing the opportunity of a formidable expansion of the nanoporous materials database triggered by the Materials Genome Initiative, Simon \emph{et al.} screened 670,000 experimental and hypothetical nanoporous material structures for Xe/Kr separation.\cite{Simon2015} It is one of the largest-scale screening performed in this area. Inspired by the work of Fernandez and co-workers,\cite{Fernandez_2013} they used ML algorithms to train a model on a diverse subset of 15,000 structures. This method allowed them to run time-consuming molecular simulations only on this training set, before applying the ML model to predict the selectivity values on the larger set of structures. On top of analysing the links between pore descriptors and selectivity, they rationalised it using theoretical pore models of spherical and cylindrical geometries to confirm the findings of Snurr and co-workers.\cite{Ryan_2010,Sikora2012} By comparing the structural descriptors of good-performing and bad-performing structures, they concluded that geometrical descriptors wasn't enough to explain the performance (see Figure \ref{fgr:Simon2015}). The analysis of a few top candidates suggests that different chemical insights could explain their good performance. For SBMOF-1 or KAXQIL,\cite{KAXQIL} an experimental MOF, its higher performance was explained by the tube-like 1D channel with a very favourable binding site formed by carbon aromatic rings. This nanoporous material was later tested using breakthrough experiments and proved to be one of the most promising candidates.\cite{Banerjee_2016} This close collaboration between computation and experimentation is a testimony of the potential of computational screenings to find nanoporous materials for any targeted application.

The experimental work on Xe/Kr separation on SBMOF-1 revealed discrepancies between the selectivity values obtained experimentally and computationally.\cite{Banerjee_2016} \revadd{The assumption of rigid crystal structures in the molecular simulations could partially explain the difference observed.} Witman \emph{et al.} proposed that the flexibility of the materials, that weren't considered in the screening of Simon \emph{et al.}, could explain the lower selectivity observed experimentally.\cite{Witman_2017} In this study, they screened the Henry regime separation of about 4,000 MOF structures of the CoRE MOF 2014 database\cite{Chung_2014}, and found that intrinsic flexibility, \emph{i.e.} the thermal vibration of the material, can make the pore size derive from the ideal value for the separation and hence lower the selectivity. This study further confirms the importance of the pore size by highlighting the effect of its evolution over time.

In 2019, Chung \emph{et al.} screened the most extensive simulation-ready and experimentally synthesised MOF structures for Xe/Kr separation.\cite{Chung2019} This study pointed out the potential of coordinated solvent molecules to fine-tune the selectivity for any separation application, since their presence can enhance selectivity in some cases. The results of their screening confirms the potential of structures such as SBMOF-1 found by Simon \emph{et al.}, but they also described a few structures with similar selectivity but with better xenon uptake. The authors emphasise the importance of considering other figures of merit such as the adsorption capacity. Other factors should be taken into account to find the best trade-off between all the relevant figures of merit; we could think of the kinetics of such a separation, the effect of flexibility on the performance, the stability of the materials (especially in radioactive environment), the financial aspects, and more. Some of these aspects will be tackled in the following sections of this review.

Beside noble gas separation, carbon capture could benefit greatly from the use of nanoporous materials and we can find extensive literature on computational screening targeting this application.\cite{Haldoupis_2012,Huck_2014,Li_2016,Darunte_2016,Park_2017,Yang_2020} Findley and Sholl performed a screening of CoRE MOF 2014 to find the best structures for CO$_2$ capture in humid conditions.\cite{Findley_2021} After finding candidates, they performed quantum calculations but found that the classical methods with generic forcefields overestimated the performance, highlighting the limits of the methodology. For a more in depth review on separation, Daglar and Kaskin described the recent development of high-throughput screening focusing mainly on CO$_2$ separation from methane of diazote.\cite{Daglar_2020}

\section{Transport properties}

In the previous section, the thermodynamic properties only described the state of equilibrium of the adsorption process. But sometimes the transient state can last long before reaching the equilibrium, which makes the process more time-consuming. Thus, the transport properties complete the thermodynamic description of the adsorption process inside a nanoporous material. For example, a low diffusion rate would mean for storage applications more time and energy needed to fill-up the tanks, or for separation applications a less selective process than expected. In more extreme cases of molecular sieves for fluid separation, the transport properties become predominant to assess the performance. One can leverage the difference of the molecules diffusion coefficients to selectively filter gas mixtures through a nanoporous membrane.\cite{Miandoab_2021} Here, the main subject becomes the transient state and not the equilibrium. This section is thus dedicated to the kinetics of the adsorption process to better model the time required to reach the equilibrium or to study out-of-equilibrium processes such as molecular sieving by nanoporous membranes.

\subsection{Diffusion calculation to model the kinetics of adsorption}

In most computational screenings, the diffusion coefficient considered is the self-diffusion coefficient that describes an infinite-dilution case. Other multi-component diffusion coefficients could be considered, but for simplicity and clarity they won't be mentioned in this review. The calculation of the self-diffusion coefficient gives a first estimation of the kinetics in a storage or a separation process in the limit of low adsorption loading.

There are two approaches to estimate the diffusion inside a porous material: the first one relies on molecular dynamics (MD) and the second one on transition state theories. In the first approach, one analyses the mean squared displacement of the adsorbed molecule moving in the material. In the second, one identifies minimum energy path along the material to identify transition states (TS) to calculate diffusion energy barriers. The MD-based method requires \revadd{fewer} assumptions and is therefore more reliable than the TS-based method, but the latter is computationally more efficient in the case of low diffusion rate (diffusivity lower than 10$^{-11}$ \si{\square\meter\per\second}).

State-of-the-art MD simulations could calculate rather accurate diffusion coefficients, but the computational cost scales quickly with the number of structures. To use this method on a large dataset without spending to much computation time, Watanabe and Sholl pre-screened the pore sizes of 1,163 MOFs to select only the structures within a certain range of PLD (pore limiting diameters).\cite{Watanabe2012} A restricted list of 359 MOFs was then used to carry out MD simulations to calculate diffusion coefficients. The results of this final screening are then used to extract the most promising structures for further experimental or computational investigation. Similarly, Qiao \emph{et al.} used a multi-stage screening to find the best membrane-material within about 130,000 hypothetical MOFs for a CO$_2$/N$_2$/CH$_4$ separation.\cite{Qiao_2016} They started to select materials based on pore geometry analysis; then they calculated Henry's coefficient and diffusion coefficients at infinite dilution; finally they compared the binary permselectivities to extract 24 promising MOFs for ternary adsorption and diffusion calculation at the desired pressure and temperature conditions.

\begin{figure}[ht]
\centering
  \includegraphics[width=0.7\linewidth]{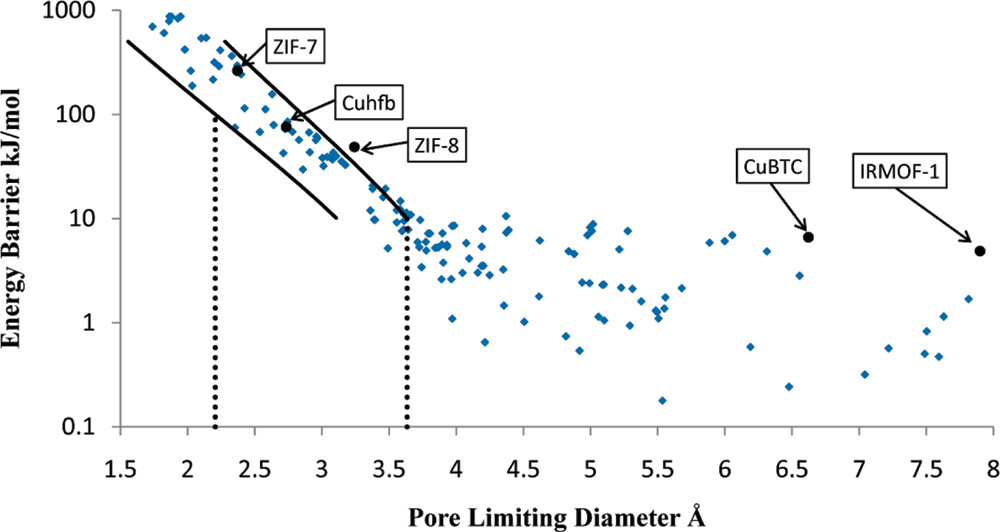}
  \caption{Calculated energy barrier for the diffusion of CH$_4$ in 216 metal--organic frameworks (MOFs), shown as a function of the pore-limiting diameter. The solid lines represents statistical upper and lower bounds on the energy barrier, in a transition state theory approach. Reprinted with permission from Ref.~\citenum{Haldoupis_2010}. Copyright 2010 American Chemical Society.}
  \label{fgr:Haldoupis_2010}
\end{figure}

Another approach replaces MD simulations with more computationally efficient TS-based methods to determine diffusion coefficients.
Haldoupis \emph{et al.} developed an algorithm to identify diffusion paths by exploiting an energy grid with a clustering algorithm. The diffusion paths are then analysed to identify the pores and the channels, and to calculate key geometric (PLD, largest cavity diameter) and energetic (Henry's constant, diffusion activation energy) features.\cite{Haldoupis_2010}
As represented in see Figure \ref{fgr:Haldoupis_2010}, they found a clear dependence of the diffusion energy barrier to the PLD. As one of the first TS-based screenings, it is still subject to many development perspectives. For instance, the approach is limited to spherical adsorbates and rigid frameworks. Moreover, the diffusion coefficients are approximated using a simplistic hopping model for a qualitative analysis. This method is highly efficient, but the accumulation of approximations makes a quantitative systematic analysis of diffusion coefficients out of reach.

Later, Kim \emph{et al.} introduced a flood fill algorithm to obtain all the points within a given energy.\cite{Kim_2013} These points are then identified as channels or blocked regions. Along the channels, local minimums of energy are defined as lattice sites and transition states are defined perpendicular to the diffusion direction. A random walk is then computed along the lattice sites with hop-rates defined according to the activation energy. A diffusion coefficient is then calculated in each three directions of the space and an average diffusion coefficients is finally determined.
A comparison with the MD method on the IZA zeolite structures shows good agreement, but there are still some discrepancies explained by correlated hops in the case of rapid diffusion or by the presence of complicated channel profiles. Inspired by this work, Mace \emph{et al.} developed a similar method that progressively fill the energy grid to detect transition states, hence removing the previous restriction to orthogonal cells only.\cite{Mace_2019} The diffusion coefficient is now computed using a kinetic Monte Carlo simulation allowing the adsorbate to jump freely in all directions instead of restricting it in a single dimension. This new method, called TuTraSt, handles very complex diffusion paths (like in the AEI zeolite). This new approach seems to be promising as it is in good agreement with MD simulations, while being 2-3 orders of magnitude faster. However, the time performance could improve tremendously by translating it from Matlab to C++ and by implementing parallelisation procedures.

Very recently a massively parallel GPU-accelerated string method has been implemented and shared publicly to compute very efficiently diffusion coefficients \revadd{based on the transition state theory}.\cite{Zhou_2021} The recent developments in the prediction of diffusion coefficients in nanoporous materials point towards a promising future for the screening of transport properties applied to even larger databases. Going further, Bukowski \emph{et al.} reviewed thoroughly diffusion in nanoporous solids as an attempt to connect theory to experiments.\cite{Bukowski_2021}

\subsection{Membrane materials for gas separation}

In separation application, the study of the transport properties can evaluate the feasibility of the thermodynamic equilibrium, crucial for any bed separation process. If this separation is not feasible, kinetic separation or partial molecular sieving are to be considered. Some notable examples are: air separation in zeolites using pressure swing adsorption,\cite{ruthven1990air} N$_2$/O$_2$ separation in carbon molecular sieves,\cite{Reid_1999} or N$_2$ removal from natural gas.\cite{Wang_2019} In kinetic separation, the valuable metric is not the selectivity anymore, but the permselectivity, \emph{i.e.} the product of the selectivity and the permeability (ratio of diffusion coefficients). Therefore, the screening of diffusion coefficients gives complementary information to the thermodynamic selectivity screenings. Here, we give some examples of such screening and the main descriptors that partially explains the computed figures of merit.

\begin{figure}[ht]
\centering
  \includegraphics[width=0.5\linewidth]{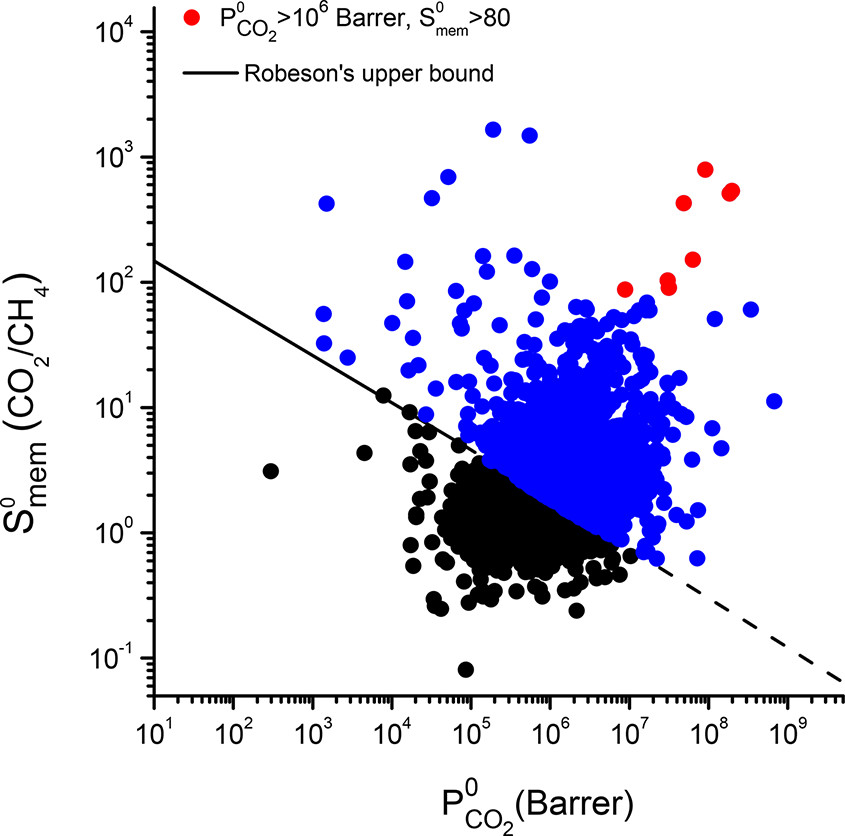}
  \caption{Selectivity and permeability of metal--organic framework (MOF) membranes for CO$_2$/CH$_4$ separation, computed at infinite dilution by combining Grand Canonical Monte Carlo and molecular dynamics simulations.\cite{Altintas_2018} The black solid line represents the Robeson's upper bound.\cite{robeson1991correlation, Robeson2008} MOFs that can exceed the bound are shown in blue, and the 8 top-performing MOF membranes are shown with red symbols. Reprinted with permission from Ref.~\citenum{Altintas_2018}. Copyright 2018 American Chemical Society.}
  \label{fgr:Altintas_2018}
\end{figure}

To give an overview on the potential of computational screenings to predict transport properties, we are now going to focus on the membrane separation applied to natural gas upgrading. The separation of CH$_4$ from N$_2$ and CO$_2$ is a crucial step of this upgrading process.
In 2016, a large scale high-throughput screening (see Figure \ref{fgr:screening} for the approach) of hypothetical MOF membranes for upgrading natural gas has been performed using MD simulations.\cite{Qiao_2016} Qiao \emph{et al.} confirmed the existence of MOF materials beyond the upper bound for N$_2$/CH$_4$ and CO$_2$/CH$_4$ separations determined by Robeson on a large set of polymeric membranes.\cite{robeson1991correlation} This Robeson's upper bound is systematically crossed by MOF materials in computational screenings, see as an example the Figure \ref{fgr:Altintas_2018}. This can be explained by the fact that MOFs perform better that polymeric frameworks and the simulations at this level of theory  They also identified 24 MOFs suitable for the ternary CO$_2$/N$_2$/CH$_4$ separation using a multi-stage screening described in the previous section.

Two years later, Qiao \emph{et al.} used the same approach to study this ternary separation on a database of synthesised structures.\cite{Qiao_2018} Applying machine learning techniques to their data, they performed a QSPR analysis. Using a principal component analysis, they notably found that the permeability is higher when materials have high PLD and void fraction coupled with low density and percentage of pores within a characteristic range. The opposite was found to be true for high membrane selectivity for the CO$_2$/CH$_4$ separation. Using decision tree algorithms, they gave objective procedures of selecting the best separation membranes based on some key descriptors. Finally, they studied in detail some of the best performing materials found by a support vector machine algorithm.

Altintas and Keskin later performed a screening on the same database for CO$_2$/CH$_4$ membrane separation to identify the best performing materials and perform more computationally demanding simulations.\cite{Altintas_2018} \revadd{The simulations in rigid structures at infinite dilution show a large number of structures above the Robeson's upper bound as shown in figure \ref{fgr:Altintas_2018}, this crossing of the upper bound can be explained by either a better performance of MOF membranes compared to the polymeric membranes used by Robeson, or an overestimation due to oversimplified assumptions (infinite dilution, rigidity). But when higher pressures and flexibility are considered, the selectivity values are dropping down closer to the upper boundary}, hence confirming the overestimation of the performance in screenings \revadd{based on rigid approximations at infinite dilution}. \revrem{But the best performing materials are still above the Robeson's upper bound and can therefore be used in mixed matrix membranes with polymeric membranes.} Budhathoki \emph{et al.} developed a screening methodology for MOFs in mixed matrix membranes for carbon capture applications by estimating permeation values in these composite materials using a Maxwell model.\cite{Budhathoki_2019} The authors even proposed a pricing for each material compared to their relative performance. Similar studies have been carried out on different materials, Yan \emph{et al.} showed the influence of decorating COFs with different chemical compounds on the membrane selectivity.\cite{Yan_2018}

The transport properties screening is based on the calculation of diffusion coefficients at infinite dilution and in rigid molecules. There are different methods to calculate them (mainly MD and TS-based methods). Flexibility and pressure dependence are very hard to incorporate directly in the screening procedures. Researchers usually consider these factors at the end of the screening on the most promising structures because of the computational complexity of the corresponding simulations. \revadd{To take account of} pressure dependence, \revadd{we need} an MD simulation of several adsorbates \revadd{ that takes much more time than running single component simulations},\cite{Keskin_2007, Keskin_2009} which \revadd{makes it harder to include in a} high-throughput screening. Flexibility could be taken account by calculating snapshots and running multiple MD simulations, or by using flexible force fields, which means in both cases an increase in computational run-time. Some faster methods of quantitatively predicting the impact of flexibility on diffusion are being investigated in ZIFs and could give an interesting alternative to these expensive methodologies.\cite{Han_2020}

\section{Non-adsorptive properties}

Due to their high internal surface area, adsorption applications were a natural outlet for nanoporous materials. However, these materials can be used in many other applications. This section is dedicated to the physical and chemical properties not directly related to the adsorption process inside nanoporous materials such as
catalytic activity,\cite{Singh2015, Greeley2006, Back2020}
mechanical properties,\cite{Chibani2019, Gaillac2020}
or thermal properties.\cite{Toher2014, Sarikurt2020, Ducamp_2021} \revadd{These properties require a more refined description of the atomic interactions within the material. DFT simulations are usually performed to accurately retrieve these properties. However, the computational cost required is multiplied by several orders of magnitude compared to classical simulations. The size of the datasets screened are therefore much smaller (a few hundreds maximum), and the use of ML can potentially speed up the whole process. ML is based on lower cost descriptors,\cite{Evans2017, Ducamp_2022} or it can be used in ML potentials for molecular simulations\cite{Eckhoff2019,Friederich2021}. }

\subsection{Catalytic activity}

Beyond adsorption properties, screening procedures have been applied to chemical properties such as catalytic activities.
Heterogeneous catalysis is generally performed using metallic nonporous structures, the use of nanoporous materials can increase dramatically the active surface area and the catalytic activity. Consequently, MOFs have been demonstrated to show catalytic properties for several chemical reactions. Just to cite a few, one can think of hydrogenation, hydrolysis, oxidation, among others explicitly covered by McCarver \emph{et al.} in their review.\cite{McCarver_2021}
Considering the sheer amount of possible materials, computational studies are potentially more effective than experimental ones. Therefore, computational screenings evolved in the last decade aiming at studying larger datasets.

Although the vast majority of computational screenings have been done on small series, there are a few systematic screenings of larger datasets. The scarcity of the latter can be explained by the high level of computational cost required. Here, we show some examples of such attempts by focusing on the example of C--H bond activation for the conversion of alkanes into alcohols in the presence of nitrous oxide.

Inspired by enzymatic catalysis of the reaction of small alkanes with N$_2$O into alcohols, Vogiatzis \emph{et al.} identified 7 iron containing MOF structures out of 5,000 structures from the CoRE MOF database.\cite{Vogiatzis_2016} They found two descriptors that govern the catalytic activity: 1) the N--O dissociation energy of N$_2$O on the adsorption site and 2) the energy difference between two spin states of the intermediate.
Using a screening on these descriptors, three structures were identified as promising for further experimental studies. The best one has been computationally demonstrated to catalytically and selectively oxidise ethane to ethanol in presence of N$_2$O. Moreover, the authors found that defects played a major role in the observed catalytic activity.

\begin{figure}[ht]
\centering
  \includegraphics[width=0.8\linewidth]{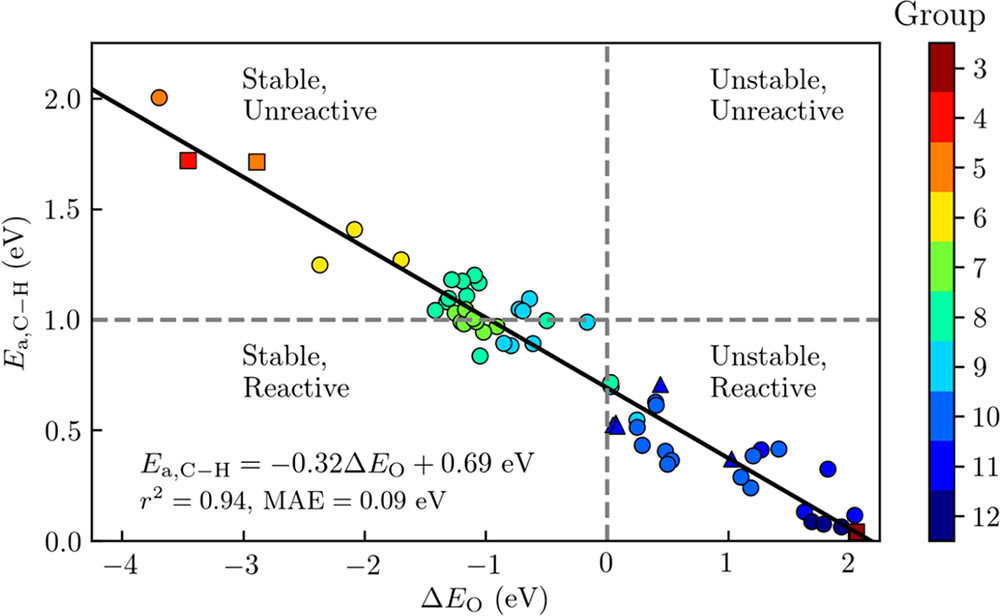}
  \caption{Analysis of a diverse set of experimentally derived metal--organic frameworks (MOFs) with accessible metal sites for the oxidative activation of methane. The graph shows the predicted barrier for the C--H bond activation of methane, E$_\text{a}$, as a function of the metal-oxo formation energy, $\Delta E_\text{O}$. For each material, the symbol colour refers to the group number of the metal in the periodic table. The best-fit line has is plotted in black, and has a mean absolute error (MAE) of \SI{0.09}{\eV}. MOFs with E$_\text{a}<$ \SI{1}{\eV} are classified as being reactive toward C--H bond activation and MOFs with $\Delta E_\text{O}<0$ as having thermodynamically favoured active sites when using O$_2$ as the reference state. Reprinted with permission from Ref.~\citenum{Rosen_2019}. Copyright 2019 American Chemical Society.}
  \label{fgr:Rosen_2019}
\end{figure}

Later, Rosen \emph{et al.} enlarged the scope of materials screened to other metals.\cite{Rosen_2019} From an 838 DFT-optimised MOFs subset of CoRE MOF 2014, the authors selected 168 MOFs that were likely to have open metal sites and pore-limiting diameters that allows the diffusion of the reactants. They then used a fully automated workflow to place the reactants in the adsorption site and relaxed the system using periodic DFT calculations. As shown in Figure \ref{fgr:Rosen_2019}, using the bond activation energy E$_\text{a,C--H}$ and the metal-oxo formation energy $\Delta E_\text{O}$ as key parameters, they classified the materials according to their relative stability and reactivity to find the best materials for the application. These energies were then analysed using physicochemical descriptors such as the spin density on the oxygen and the metal--oxygen distance.

This type of brute force screening can be quickly cumbersome, as a result many researchers in the field are trying to find key structure-activity relationships to accelerate future computational screenings.
Several descriptors have been developed for high-throughput screenings: Butler \emph{et al.} used electron removal energies to explain photocatalytic behaviours of MOFs;\cite{Butler_2014} Rosen \emph{et al.} showed that the energy required to form the metal-oxide intermediate was a key descriptor of the thermal catalysis of alkane oxidation by N$_2$O;\cite{Rosen_HTPDFT_2019} and Fumanal \emph{et al.} show a screening protocol based on two energy-based descriptors to predict photocatalytic properties of MOFs.\cite{Fumanal_descriptor_2020} \revadd{Lately, Rosen \emph{et al.} screened thousands of MOF structures to compare different DFT functionals and leveraged the data calculated to train machine learning models that can rapidly predict MOF band gaps.\cite{Rosen2022} }

The development of ML methods are also critical in the field,\cite{Rosen_2021} but the lack of centralised database with high precision descriptors is a challenge for the future of these methods. The influence of defects, the different ways of modelling MOFs as periodic structures or clusters, the diversity of structures and the stability of such structures remain open problems. Yet, it does not threaten the major role of high-throughput screenings in the early design process of any nanoporous materials for catalysis. To conclude this brief overview, we point the readers to a more exhaustive presentation of the matter.\cite{Rosen_2022}

\subsection{Mechanical properties}

In the past decade, there has been a growing interest in the systematic study of physical properties of various classes of materials, including inorganic materials and framework materials. Among these physical properties, mechanical properties have been a topic of particular interest, as they are crucial for many applications, and at the same time can be computed by relatively standard methodologies. In particular, is it possible to calculate linear elastic constants (the second-order elastic tensor) in the zero-Kelvin limit by strain/stress or strain/energy approaches, performing a series of DFT calculations of strained structures and calculating the elastic constants. From these constants, all other mechanical properties can be evaluated by tensorial analysis,\cite{Marmier2010} including the bulk modulus, Young's modulus, shear modulus, Poisson's ratio, etc. This type of calculation can be coupled with any available quantum chemistry code,\cite{Golesorkhtabar2013} and is even integrated in some packages, like CRYSTAL17.\cite{Dovesi2018}

One of the first studies that investigated systematically the elastic properties of a family of materials was a 2013 study of all-silica zeolites,\cite{Coudert2013} i.e., crystalline and porous SiO$_2$ polymorphs. While this dealt with only 121 zeolitic frameworks out of 244 known structures, it showed that systematic studies at the DFT level were computationally tractable, and that they provided physical insight into the link between microscopic structure and macroscopic physical properties. This study demonstrated, among other things, that a small number of zeolites presented large negative linear compressibility (NLC), which could be linked to the wine-rack motif of their frameworks.

\begin{figure}[ht]
\centering
  \includegraphics[width=0.8\linewidth]{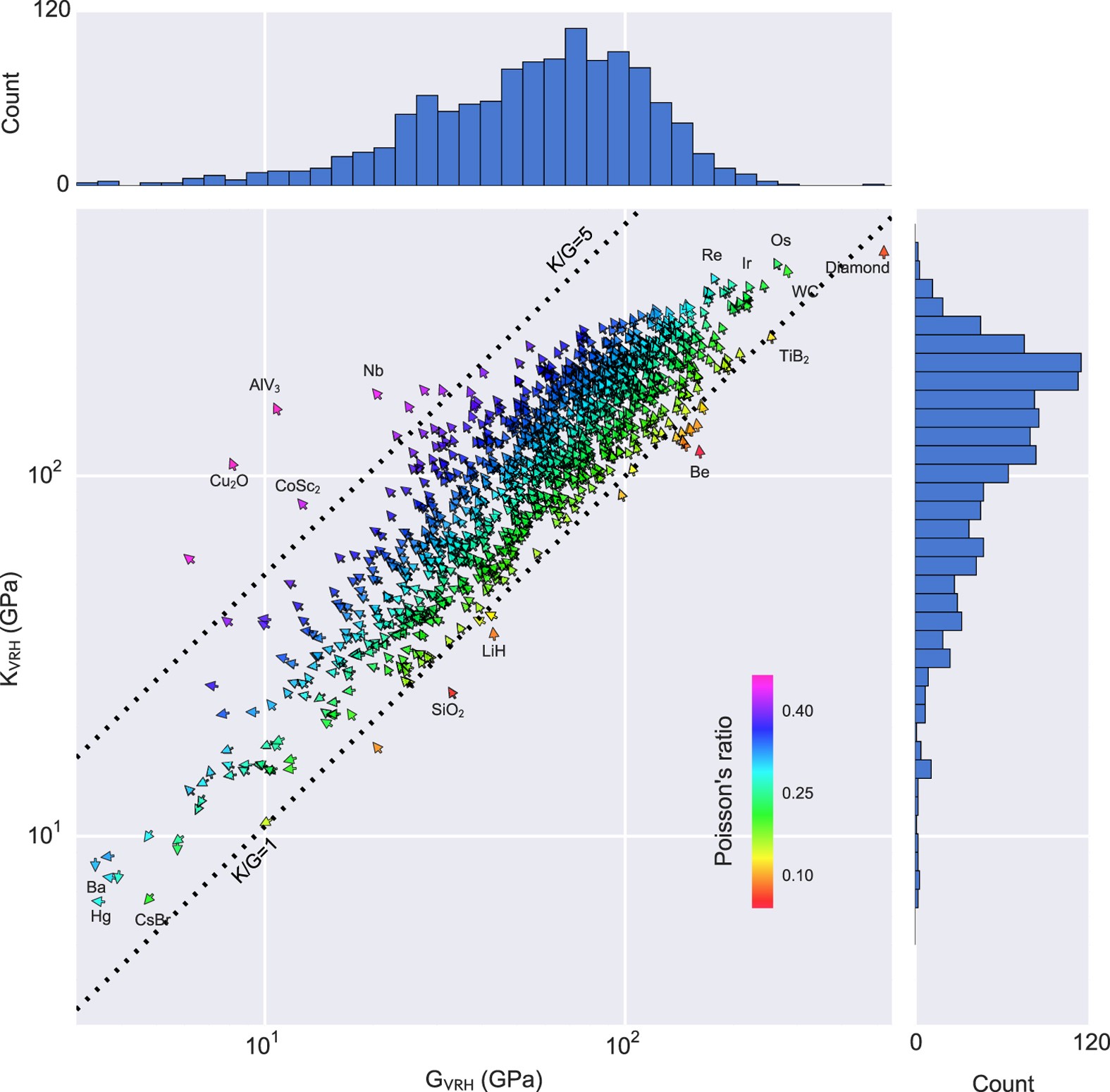}
  \caption{Statistical analysis of the calculated volume per atom, Poisson's ratio, bulk modulus \revadd{$\mathrm{K_{VRH}}$} and shear modulus \revadd{$\mathrm{G_{VRH}}$ of} 1,181 compounds in the Materials Project database. In the vector field-plot, arrows pointing at 12 o'clock correspond to minimum volume-per-atom and move anti-clockwise in the direction of maximum volume-per-atom, which is located at 6 o'clock. Reprinted from Ref.~\citenum{deJong2015} under CC-BY license. Copyright 2015 de Jong \emph{et al}.}
  \label{fgr:deJong2015}
\end{figure}

Looking outside of the specific case of zeolites, other groups have applied DFT calculations of elastic constants in a high-throughput manner. de Jong \emph{et al.} leveraged the structures of the Materials Project\cite{Mat_genome, Jain_2013}, trying to chart the diversity of elastic properties across the whole space of inorganic crystalline compounds.\cite{deJong2015} As shown in the Figure \ref{fgr:deJong2015}, they provided a database containing the full elastic information of 1,181 inorganic compounds initially, and has grown steadily since then, containing more almost 14,000 records to date.\cite{MaterialsProject} This dataset has been used in two different ways by researchers in the field.

Firstly, the exploration of the database of elastic properties by tensorial analysis has allowed to study quantitatively the occurrence of certain ``anomalous'' or rare mechanical behaviour, including negative linear compressibility, very high anisotropy, or negative Poisson's ratio (also called \emph{auxeticity}). Indeed, such properties are considered rare and usually sought after --- the materials exhibiting these anomalous behaviours are mechanical metamaterials.\cite{Coudert2019_meta} \revadd{In addition to their fundamental interest, such materials have applications in materials engineering: for example in energy dissipation (as shock absorbers and for bulletproofing), energy storage, as well as acoustics.\cite{Surjadi2018}} However, it was not possible until now to quantify exactly ``how rare'' they are. Chibani \emph{et al.} showed through a systematic exploration of available mechanical properties of crystalline materials that general mechanical trends, which hold for isotropic (noncrystalline) materials at the macroscopic scale, also apply on average for crystals. Moreover, they could quantify the presence of materials with rare anomalous mechanical properties: {3\%} of the crystals were found to feature negative linear compressibility, and only {0.3\%} to exhibit complete auxeticity (negative Poisson's ratio in all directions of space).

Secondly, the datasets of mechanical properties were used as a basis to accelerate the discovery of novel materials with targeted behaviour. Dagdelen \emph{et al.} used search algorithms to identify 38 candidate materials exhibiting features correlating with auxetic behaviour, from more than 67,000 materials in the Materials Project database.\cite{Dagdelen2017} Performing DFT calculations on these 38 structures, they could identify 7 new auxetic compounds. In a more complex setup, Gaillac \emph{et al.} \cite{Gaillac2020} have used a multi-scale modelling strategy for the fast exploration and identification of novel auxetic materials. They combined classical force fields MD simulations with DFT calculations on candidate materials, and then used this reference DFT data to train an ML algorithm. They found that the accuracy of this multi-scale method exceeds the current low-computational-cost approaches for screening. In a similar work, Moghadam \emph{et al.} used molecular simulation to train an artificial neural network (ANN) for the prediction of the bulk modulus of metal--organic frameworks.\cite{Moghadam_2019} This shows the potential of such methodologies to treat very different (chemically as well as structurally) classes of materials.

\subsection{Thermal properties}

While mechanical properties (in the elastic regime) have been by far the most studied physical property in nanoporous materials, others have also been occasionally screened. We can cite, in particular, the systematic study of piezoelectric tensors by de Jong \emph{et al.}, on almost a thousand crystalline compounds, by first-principles calculations based on density functional perturbation theory.\cite{deJong2015_piezo} We can also cite efforts to calculate thermal properties in a high-throughput setup, using the quasi-harmonic approximation (QHA).\cite{Togo2010} This method requires the calculation of each structure's phonon modes at various volumes, and can be coupled to any electronic structure program.\cite{Togo2015} It is, however, quite computationally intensive, and sensitive to the parameters of the QHA methodology (range of volume, range of temperature, precision of the frequency calculation, etc.). Therefore, it has been limited so far to modest numbers of structures: a dataset of 75 inorganic structures by Toher \emph{et al.},\cite{Toher2014} and more recently a dataset of 134 pure SiO$_2$ zeolites by Ducamp \emph{et al.}\cite{Ducamp_2021} \revadd{Very recent work in our group on the prediction of thermal properties through machine learning based on structural features alone indicates that thermal behaviour is more difficult than mechanical behaviour to predict, and might require the use of a wider set of structural descriptors, or more advanced ML models.\cite{Ducamp_2022}}

\section{Outlook}

In this review, we highlighted the advances in computational screening of nanoporous materials for some archetypal cases through the scope of their physical and chemical properties. Although each type of property requires a specific simulation methodology and has distinct challenges, the essence and general workflow of high-throughput screening does not fundamentally change. The goal is to generate quickly and accurately increasing amounts of valuable data in order to analyse it. With the increase of high-performance computing (HPC) resources and the help of statistical tools such as machine learning, screening techniques have seen a rapid acceleration in recent years. Researchers can more efficiently analyse larger and larger databases and help theoreticians better understand the origins of the performance, hence guiding the design process of nanoporous materials.

\revadd{
Despite the progress made, important drawbacks of the current methodologies remain. High-throughput screenings rely too much on oversimplified assumptions such as the rigidity of the framework, the absence of defects, the use of Lennard-Jones potentials and inaccurate charges. For instance, the rigidity of the framework only takes into account one conformation of the framework. Yet, thermal agitation induces a ``breathing'' movement of the framework with an amplitude dependent on its intrinsic flexibility. The pores of the framework can change depending on the number of adsorbates to interact more optimally with them, which can be induced by a change in pressure. The issue of flexibility is rarely tackled, and when considered, it is only on the few most selective structures given by an inaccurate screening based on the rigid crystal approximation. One can wonder about the results obtained if it is applied to larger sets of structures. Witman \emph{et al.} found that flexibility applied to top performing materials can decrease the selectivity, because the pore does not have an optimal size anymore.\cite{Witman_2017} In some cases, the selectivity of a well performing material can even increase to become a top performing one. Computational screenings can be closer to predict experimental values of selectivity, diffusivity, and other key performance metrics. }

\revadd{Many open problems remain for the design of efficient high-throughput computational screenings}. The connection between different properties for a given application is not systematically integrated in the screening procedures. For example, in methane storage, the working capacity of the material is the main property to optimise, but the kinetics of the adsorption/desorption or the mechanical resistance to compaction amongst others also need to be considered. Designing a nanoporous material is in fact a multivariate optimisation problem with tacit constrain\revadd{t}s, for example the \revadd{synthesisability}. Moreover, \revadd{the transferability of the methodology to a broad range of materials is often achieved at the expense of accuracy in specific cases.} \revadd{And one can rightly question the universality of depending on faster but less elaborated models}, which boils down to a trade-off problem between prediction accuracy and computational cost (or complexity). For instance, classical force-fields are broadly used in rigid materials for adsorption properties, but the switch to more costly \emph{ab initio} methods or the addition of flexibility can result in a more accurate description at the expense of computational resources. The use of ML algorithms can be a way out of this apparent deadlock. They can learn sufficient information on as small a subset as possible to accurately predict the performance of other materials on a large dataset. It could in the future reduce the size of the dataset that needs to be accurately screened by computationally expensive simulations, while maintaining the quality of the predictions.

The development of such ML-assisted screenings is paired with the advances in data science techniques and algorithms, but more importantly to the construction of descriptors tailored to the many possible application. This construction work cannot be dissociated to the physical and chemical intuition of the scientists. Topological, chemical, electronic and other descriptors have been developed on top of the more common geometrical and thermodynamic descriptors, which displays the importance of strong physical chemistry knowledge. The discovery of novel relevant descriptors remains the main lever for increased performance of the ML models and is closely related to a rigorous theoretical work.

The development of databases is another key aspect in the promotion of data science in the field of materials science in general, and nanoporous materials chemistry in particular. The diversity of materials, the inclusion of experimental data (successful or failed), the addition of under studied classes of materials (\emph{e.g.} amorphous) are all key aspects to upgrade the existing database. Even if existing attempts to create a centralised database have been initiated by the materials project,\cite{MaterialsProject} this database does not contain all the existing information on each material.

In the future, computational high-throughput screening could be integrated more tightly into the design process of nanoporous materials, hence further improving its efficiency. The computational pre-screening can be coupled with automated screenings of the most promising materials to finally identify candidates for further studies. This automated design process is described by Lyu \emph{et al.} in their paper on ``Digital Reticular Chemistry'' and set out promising perspectives for computational screenings in the field.\cite{Lyu_2020} \revadd{Some studies are already pioneering this new research area by combining high-throughput characterisations, active learning algorithms and robotic synthesis.\cite{Greenaway2018,Moosavi2019} Another step towards faster industrialisation would integrate process modelling to enrich the purely atomistic approach.}

\section*{Acknowledgements}
We thank Isabelle Hablot for discussions on the topic of adsorption-based separation. This work was posted as a preprint at \url{https://arxiv.org/abs/2202.09886}

\section*{Funding}
This work was financially supported by Orano.


\bibliographystyle{tfnlm}
\bibliography{article}

\end{document}